\documentclass[aps,prb,twocolumn,superscriptaddress,showpacs,amsmath]{revtex4}
\usepackage{amsfonts}
\usepackage{amssymb,latexsym}
\usepackage{graphicx}
\newcommand{\be}{\begin{equation}}
\newcommand{\ee}{\end{equation}}
\newcommand{\bea}{\begin{eqnarray}}
\newcommand{\eea}{\end{eqnarray}}

\begin{document}

\title{Quantum dimer phases in a frustrated spin ladder:
Effective field theory approach and exact diagonalization}

\author{Temo Vekua}
\affiliation{Universit\'{e} Louis Pasteur,
  Laboratoire de Physique Th\'{e}orique,
  67084 Strasbourg, France}
\affiliation{Andronikashvili Institute of Physics, Tamarashvili 6, 0177 Tbilisi, Georgia}

\author{Andreas Honecker}
\affiliation{TU Braunschweig, Institut f\"ur Theoretische Physik,
  38106 Braunschweig, Germany}
\affiliation{Institut f\"ur Theoretische Physik, Universit\"at G\"ottingen,
37077 G\"ottingen, Germany}

\begin{abstract}
The phase diagram of a frustrated $S=1/2$ antiferromagnetic spin ladder
with additional next-nearest neighbor exchanges, both diagonal and inchain,
is studied by a weak-coupling effective field theory approach
combined with exact diagonalization for finite systems.
In addition to two known phases with rung-singlet
and Haldane-type ground states, we observe two new
phases with dimerization along the chains. Furthermore, the transitions
between the different phases are studied and shown to be either first
order or to belong to the universality class of the two-dimensional
Ising model. The nature of elementary excitations is discussed briefly.
\end{abstract}

\date{January 24, 2006; revised April 25, 2006}

\pacs{75.10.Pq, 75.40.Cx, 75.30.Kz, 75.40.Mg}

\maketitle

\section{Introduction}
During the past few decades, strongly correlated electron systems have
received a lot of attention, e.g.\ due to unconventional superconductors.
In particular, during the early days of high-$T_c$
superconductivity, Anderson had already proposed a mechanism
based on a so-called `resonating valence-bond' (RVB) picture.\cite{Anderson}
Nevertheless, the search for a spin liquid in two dimensions
remains a long standing problem. In this context,
much attention was devoted to the study of geometrically
frustrated magnetic systems (for recent reviews see e.g.\
Refs.~\onlinecite{Misguich,RSH04}).

Among the different proposed models to capture Anderson's RVB scenario
is the one of Nersesyan and Tsvelik,\cite{NersTsvelik} based on
a spatially anisotropic $J_1-J_2$ square lattice with
stronger exchanges along one particular `chain' direction to allow for
generalizing results obtained previously in one dimension.\cite{Allen}
However, it turned out that
even the one-dimensional picture was not completely under control:\cite{Balents}
in weak coupling it is impossible to
fine-tune rung and diagonal exchanges of two-leg spin ladders to
eliminate all relevant interchain couplings at all orders.
In particular it was argued in Ref.~\onlinecite{Balents} that
at weak coupling there is no direct first-order phase
transition from two phases with unique ground states,
namely a Haldane phase and a rung-singlet phase, rather an
intermediate, spontaneously dimerized phase was predicted.
However, previous numerical calculations have not detected
this intermediate phase.\cite{Wang} The numerically determined
phase diagram suggested a direct phase transition from
the Haldane to the rung-singlet phase which at weak interchain coupling
seemed to be of second order and become first order at stronger
couplings. While the latter topology of the phase diagram agrees
with the earlier bosonization prediction,\cite{Allen} the
order of the phase transitions disagrees at weak coupling.
We will argue that the intermediate phase in the spin ladder
can be revealed by adding further exchanges.

In this work we study an $S=1/2$ antiferromagnetic two-leg spin ladder
with additional next-nearest neighbor exchanges and discuss
the interplay between two distinct scales, one set by inchain
frustration and the other one by interchain interactions.
In the spirit of Ref.~\onlinecite{Capriotti},
this model can also be regarded as a strip of the extensively
studied $J_1-J_2-J_3$ square lattice.\cite{GSHLL,SachdevBhatt}
The latter model illustrates that frustrated interactions in
quantum spin systems give rise to `unpredictable' phases which
are very difficult to analyze theoretically because of the presence
of competing interactions. The study of the effect of additional
next-nearest neighbor exchanges in spin ladder systems therefore
also is an interesting theoretical problem in its own right.
The bosonization approach\cite{GNT} is an unbiased and  powerful method for
studying frustrated systems in particular in one dimension.
Using this method, we will show that a rich phase diagram and
intriguing nature of elementary excitations result from the competing
interactions in the frustrated spin ladder. Furthermore,
numerical analysis of finite systems confirms the presence of the
phases obtained within the weak-coupling effective field theory,
also at larger couplings.

\section{Model}

\begin{figure}[tb!]
\centering
\includegraphics[width=70mm]{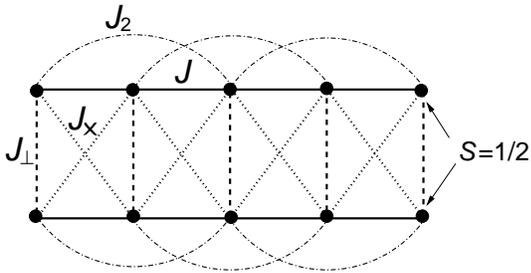}
\caption{\label{fig:ld}Structure of the spin ladder with
next-nearest neighbor interactions.}
\end{figure}

We consider the $S=1/2$ antiferromagnetic Heisenberg spin ladder with
additional next-nearest neighbor exchanges which are represented by
ladder diagonal and inchain next-nearest neighbor interactions. The
geometry of our model is depicted in Fig.\ \ref{fig:ld}.
The lattice Hamiltonian reads:
\begin{equation}
\label{LatticeHamiltonian}
\hat{H} = H_{leg}^{1} + H_{leg}^{2} +  H_{int}\, ,
\end{equation}
where the Hamiltonian for leg $l=1,2$ is
\bea\label{InchainHamiltonian}
H_{leg}^{l} &=&\sum_{j=1}^{N}
\left(J \, {\mathbf S}_{l,j}\cdot{\mathbf S}_{l,j+1} +
J_2 \, {\mathbf S}_{l,j}\cdot{\mathbf S}_{l,j+2}\right)
\eea
and the interleg coupling contains both the rung and the diagonal exchanges:
\bea
\label{InterLegCoupling}
H_{int} = \sum_{j=1}^{N} &&\left[
 J_{\bot} \, {\mathbf S}_{1,j}\cdot{\mathbf S}_{2,j} \right.  \\
&& \left.+J_{\times} \, \left( {\mathbf S}_{1,j}\cdot{\mathbf S}_{2,j+1}+
{\mathbf S}_{1,j+1}\cdot{\mathbf S}_{2,j} \right)\right] \, . \nonumber
\eea
Above ${\mathbf S}_{l,j}$ represent spin $S=1/2$ operators on the $j$-th
 rung, $l$-th leg, and periodic boundary conditions are assumed
${\mathbf S}_{l,N+1}={\mathbf S}_{l,1}$. In order to avoid
additional frustration by the boundary conditions, we will consider
only even chain lengths $N$.

The above Hamiltonian (\ref{LatticeHamiltonian}) with $J_2=0$
has already been investigated in detail.\cite{Allen,Wang,Gelfand,Oitmaa}
In particular the case $J_\times = J$, $J_2 = 0$ serves as a good illustration
of two possible phases\cite{Gelfand} since in this case the total spins on
all rungs are good quantum numbers.
Then the ground state consists of local singlets
$S_{\rm rung} = 0$ for sufficiently large $J_\bot$ with a clear gap to
all excitations. On the other hand, for small $J_\bot$, total spin
$S_{\rm rung} = 1$ is formed on each rung and the low-energy physics
is governed by an effective
spin-one chain. In the latter case, the ground state is the famous
`Haldane' state which is also gapped (see e.g.\ Ref.\ \onlinecite{KoMi}
for a recent review of spin-chain models). These two phases extend
to $J_\times \ne J$, $J_2 \ne 0$ and we refer to them
as `rung-singlet' and `Haldane' phase, respectively.


\section{Weak-coupling approach, bosonization}

\label{secWeak}

In this section we perform the weak-coupling analysis of our
system (\ref{LatticeHamiltonian}).
We will follow the usual way and start from a continuous field theory description of
the individual spin chain and treat the interchain interactions perturbatively.\cite{GNT}
We note that one can construct two equivalent weak-coupling formulations.
One can start from decoupled chain limits where chains run either along the ladder
legs, or along the ladder diagonals. These two cases are connected
to each other by a `duality transformation' exchanging the spins on
every second rung.\cite{Oitmaa}
For definiteness we start from two decoupled frustrated chains running along the
ladder legs: $J_{\times},\, J_{\bot}\ll J$.
Spin operators on each chain are decomposed in their smooth and staggered parts:
\begin{equation}
 \frac{{\mathbf S}_{l,j}}{a_0}=  {\mathbf S}_{l}(x=ja_0)={\mathbf J}_{l}(x)+\bar{\mathbf J}_{l}(x)+(-1)^x {\mathbf n}_{l}(x)
\end{equation}
with lattice constant $a_0$. ${\mathbf J}_{l}$ and $ \bar{\mathbf J}_{l}$ are left and right $SU(2)$ currents of the $l$-th chain and have
conformal weights (1,0) and (0,1) respectively. Staggered spin densities are represented by more relevant operators,
indicating an inherent instability towards doubling of the unit cell
in antiferromagnetic chains:
\begin{equation}
n^i_{l}(x)\sim {\rm Tr}\big( \sigma^ig_l(x)\big) \, .
\end{equation}
Here $g_l(x)$ stands for the basic $2\times2$ matrix field of the
Wess-Zumino model and has conformal weights (1/4,1/4).
${\sigma^i}$ are the Pauli matrices.
The scalar quantity which is represented in the continuum limit by
an operator of the same dimension as the staggered part of
the magnetization is the dimerization operator:
\begin{equation}
{\mathbf S}_{l}(x){\mathbf S}_{l}(x+a_0)=(-1)^x \epsilon_{l}(x)+
\mbox{less relevant smooth part.}
\end{equation}
In terms of the basic Wess-Zumino matrix field the following representation
holds for the dimerization operator:
\begin{equation}
\epsilon_{l}(x) \sim {\rm Tr} g_l(x) \, .
\end{equation}
The effective quantum field theory in the continuum limit of a single chain is that of a critical
$SU(2)_1$ Wess-Zumino model perturbed by a marginal
current-current interaction:\cite{Affleck,CaPu}
\begin{equation}
\label{WZ}
H_{l} = \frac{2\pi u}{3} \left(:{\mathbf J}_{l} {\mathbf J}_{l}:
 +:\bar{\mathbf J}_{l}\bar{\mathbf J}_{l}:\right)
 + \gamma \, {\mathbf J}_{l}\bar{\mathbf J}_{l}
\end{equation}
with the following notations:
$$u=\frac{Ja_0\pi}{2}\qquad\mathrm{and}\qquad \gamma=J_2-J_{2,c} \, ,$$
where $J_{2,c}\simeq 0.24$.\cite{Nomura,Eggert}
If $\gamma<0$ the perturbation is marginally irrelevant,
but when $\gamma>0$ the interaction flows towards strong coupling and
the system dimerizes spontaneously with a dynamically generated gap
in the excitation spectrum.\cite{Haldane}
In Abelian bosonization representation we can rewrite (\ref{WZ}) as:
\begin{eqnarray}
H_{l}&=&\frac{u}{2}\left[(\theta_{l}(x))^2+(\partial_x\phi_{l}(x))^2 \right]\nonumber\\
&&-\frac{\gamma}{2\pi^2}\cos\sqrt{8\pi}\phi_{l}+\frac{\gamma}{\pi}\partial_x\phi_{l,R}\partial_x\phi_{l,L}
\end{eqnarray}
with compactified dual bosonic fields $\theta$ and $\phi$.
The following representations hold for
the staggered spin density and dimerization operators:\cite{GNT}
\begin{eqnarray}
{\mathbf n_l(x)}&\sim&\left(\cos \sqrt{2\pi}\theta_l(x), \sin
\sqrt{2\pi}\theta_l(x), \sin \sqrt{2\pi}\phi_l(x)\right)\,  ,\nonumber\\
\epsilon_l(x) &\sim& \cos\sqrt{2\pi}\phi_l(x) \, .
\end{eqnarray}
For further analyses, treating interchain coupling perturbatively,
it will be convenient to pass to symmetric and antisymmetric combinations of the bosonic fields
\begin{equation}
\phi_{\pm}(x)=\frac{\phi_1(x)\pm \phi_2(x)}{\sqrt {2}}
\end{equation}
and define the continuum limit expressions of order parameters for
columnar and staggered dimerizations of the two chains:
\begin{eqnarray}
\epsilon_+&=&\epsilon_1+\epsilon_2\sim \cos \sqrt{\pi}\phi_+ \cos
\sqrt{\pi}\phi_- \, ,\nonumber\\
\epsilon_-&=&\epsilon_1-\epsilon_2\sim \sin \sqrt{\pi}\phi_+ \sin
\sqrt{\pi}\phi_- \, .
\end{eqnarray}
The Hamiltonian in the symmetric and antisymmetric sectors will
contain a marginally relevant intersector coupling which promotes
dimerization within the chains:
\begin{eqnarray}
\label{Hdim}
 \frac{\gamma}{\pi}\partial_x\phi_{+,R}\partial_x\phi_{+,L} &+& \frac{\gamma}{\pi}\partial_x\phi_{-,R}\partial_x\phi_{-,L} \nonumber\\
&-& \frac{\gamma}{\pi^2} \cos \sqrt{4\pi}\phi_+ \cos\sqrt{4\pi}\phi_- \, .
\end{eqnarray}

The intersector interactions (\ref{Hdim}) show that the
vacuum configurations of the symmetric and antisymmetric sectors are
degenerate:
\begin{eqnarray}
\label{fourfolddegeneracy}
&{\rm a:}& \quad \left\langle \cos \sqrt{4\pi}\phi_+  \right\rangle=\left\langle \cos \sqrt{4\pi}\phi_-  \right\rangle=1\nonumber\\
&{\rm b:}& \quad\left\langle \cos \sqrt{4\pi}\phi_+  \right\rangle
= \left\langle \cos \sqrt{4\pi}\phi_-  \right\rangle=-1 \, .
\end{eqnarray}
This includes the two-leg ladder dimerized in different patterns,
namely a: with long-range ordered columnar
b: staggered dimerizations, respectively.

The expressions of spin operators in terms of the bosonic fields are
valid as long as the dimerization is very weak.
We can then add interchain interactions to the frustrated decoupled chains.
We classify interchain couplings according to their scaling
dimension:
\begin{eqnarray}
 H_{int} &\sim& (J_{\bot}+2\,J_{\times})\left({\mathbf J}_1(x)
+ \bar{\mathbf J}_1(x)\right) \left( {\mathbf J}_2(x)+\bar{\mathbf J}_2(x)\right)
\nonumber\\
&+&(J_{\bot}-2\,J_{\times}) \, {\mathbf n}_1(x){\mathbf n}_2(x) \, .
\end{eqnarray}
The product of the smooth parts of the spin operators is translated
into marginal operators, whereas the product of the staggered parts
is represented by relevant operators in the effective field theory.

\subsection{Mean-field separation}

As a first approximation and for $\gamma=0$, $J_{\bot}-2\,J_{\times}\ne 0$,
one can retain only relevant terms that stem
from the product of the N\'eel components of the spin operators.
In this case the effective field-theoretic Hamiltonian separates into
two commuting parts in the symmetric and antisymmetric basis:\cite{KS95,Shelton}
\begin{eqnarray}
\label{symantisym}
H & = &H^{+}+H^{-}\nonumber\\
H^{+} & =& {u_{+} \over 2} [(\partial_x \theta_{+})^{2} + (\partial_x \phi_{+})^2] - \frac{\tilde{J}_{\bot}c^2}{2\pi^2} \cos{\sqrt {4 \pi}\phi_{+}(x)} \nonumber\\
{H}^{-} & = &  {u_{-} \over 2}[(\partial_x \theta _{-})^{2}
+ (\partial_x \phi_{-}(x))^2] \\
&+& \frac {\tilde{J}_{\bot}c^2}{2\pi ^2} \cos{ \sqrt {4\pi}\phi_{-}(x)}
+ \frac{\tilde{J}_{\bot}c^2}{\pi ^2} \cos{\sqrt{4 \pi} \theta_{-}(x)} \, ,
\nonumber
\end{eqnarray}
where $\tilde{J}_{\bot}=J_{\bot}-2\,J_{\times}$, $c$ stands
for a non-universal numerical constant and
$$
u_{\pm}=u\left(1\pm\frac{\tilde{J}_{\bot}}{2\pi J}\right)
$$
in weak coupling.

Now we add $\gamma>0$ and look at the limit
$J_{\bot},J_{\times}\ll \gamma \ll J$, which allows to assume
stability of the massive dimerized phases of the individual chains against
infinetisimal interchain perturbation. For ${J}_{\bot}= J_{\times}=0$
the chains dimerize in columnar and staggered patterns which corresponds
to pinning of the symmetric and antisymmetric fields in
degenerate vacua with non-zero averages of $\cos \sqrt{4\pi}\phi_{\pm}$.
One may assume that these averages will be non-zero in a finite region of
the parameter space after inclusion of interchain coupling such that we
may substitute finite averages for $\cos \sqrt{4\pi}\phi_{\pm}$ in (\ref{Hdim}).
In the antisymmetric sector
the Hamiltonian density contains a relevant $cosine$ of the bosonic
field $\phi_{-}$ as well as its dual $\theta_{-}$, see (\ref{symantisym}).
By contrast, the dual field cannot appear
in the symmetric sector for symmetry reasons. The classical vacuum
configuration of the symmetric field is pinned unambiguously according
to the sign of interchain coupling and thus any small 
interchain coupling (which is relevant) removes immediately
the degeneracy of the ground state of decoupled
dimerized chains (\ref{fourfolddegeneracy}).
The picture is as follows: addition of infinitesimally small relevant
interchain coupling confines the massive spinons of the individual
chains into magnons. At the same time, the ground-state degeneracy is
lifted and one of the long-range ordered dimerization patterns
(columnar or staggered) is selected.
Once the symmetric field $\phi_+$ is pinned, we can perform a mean-field
decoupling of the interaction term stemming from $\gamma$.
Note that this procedure does not depend on the sign of
$J_{\bot}-2\,J_{\times}\neq 0$, the crucial assumption is that a relevant
interchain coupling is present, which provides a confining potential
to spinons of the individual dimerized chains and selects one of the two
long-range ordered dimerization patterns.

At the mean-field level we are left with the following Hamiltonian:
\begin{eqnarray}
\label{Meanfield}
H_{MF}&=&H_0+\frac{\tilde{J}_{\bot}c^2}{\pi ^2} \cos{\sqrt{4 \pi} \theta_{-}(x)}\nonumber\\
&-& \left(\frac{\tilde{J}_{\bot}c^2}{2\pi^2} +\frac{\gamma}{\pi^2}\left\langle \cos{\sqrt {4 \pi}\phi_{-}(x)}\right\rangle \right) \cos{\sqrt {4 \pi}\phi_{+}(x)} \nonumber\\
&+& \left(\frac {\tilde{J}_{\bot}c^2}{2\pi ^2}
 - \frac{\gamma}{\pi^2} \left\langle \cos{\sqrt {4 \pi}\phi_{+}(x)}\right\rangle \right)
   \cos{ \sqrt {4\pi}\phi_{-}(x)}\, ,\nonumber\\
\end{eqnarray}
where $H_0$ stands for the sum of the Gaussian parts of the symmetric
and antisymmetric sectors.
This particular form of the mean-field Hamiltonian is convenient for passing to
Majorana fermions, but let us first discuss artifacts of the
mean-field separation:
the mean-field separation in (\ref{Meanfield}) breaks the underlying
$SU(2)$ symmetry of the ladder model to $U(1)\otimes {\mathbb Z}^2$.
This is verified directly by noting that the very form of the
mean-field Hamiltonian (\ref{Meanfield}) coincides with the Hamiltonian
of a ladder where the next-nearest inchain frustrating coupling is
substituted by the product of dimerization operators of the two chains
and interchain coupling is of easy-plane type:
\begin{equation}
(A+B)\epsilon_1(x)\epsilon_2(x)+(A-B)n_1^zn_2^z
\end{equation}
with
\begin{eqnarray*}
A&=&-\frac{\gamma}{\pi^2}  \left\langle \cos{\sqrt {4 \pi}\phi_{+}(x)}\right\rangle \, , \nonumber\\
B&=&-\frac{\gamma}{\pi^2}  \left\langle \cos{\sqrt {4 \pi}\phi_{-}(x)}\right\rangle \, .
\end{eqnarray*}
While the product of dimerization operators respects $SU(2)$ symmetry,
the appearance of an effective anisotropy in interchain coupling is
clearly an artifact of the mean-field separation and should not be
taken physically (e.g.\ splitting of
the triplet of Majorana fermions into doublet and singlet).
With increasing interchain exchange in (\ref{Meanfield}) we observe a phase transition
represented by a self-dual point in the antisymmetric sector:
\begin{equation}
\frac {\tilde{J}_{\bot}c^2}{2\pi ^2}- \frac{\gamma}{\pi^2} \left\langle \cos{\sqrt {4
\pi}\phi_{+}(x)}\right\rangle=-\frac{\tilde{J}_{\bot}c^2}{\pi ^2} \, .
\end{equation}

Since our approach is a weak coupling one (starting from decoupled chains) 
we are dealing with a fixed-point Hamiltonian which is described in terms of 
two copies of the $SU(2)_1$ Wess-Zumino-Witten (WZW) model
that is equivalent to an $SO(4)$ WZW model. The $SO(4)$ WZW model
admits a representation of its generators in terms of four real
Majorana fermions.\cite{Allen97}
To clarify the symmetries of the stabilized dimerization pattern and to
unveil the nature of the self-dual point in the antisymmetric sector we
pass to these Majorana fermions and subsequently to Ising variables.

The mean-field Hamiltonian (\ref{Meanfield}) takes the following form
in Majorana fermions:\cite{Shelton}
\begin{eqnarray}\label{eq:mass}
H&=&- \int dx \left[\frac{iv_s}{2}(\rho_R \partial_x \rho_R
-\rho_L \partial_x \rho_L ) + im_s\rho_R \rho_L \right]\nonumber\\
&-&\sum_{a=1}^3 \int dx \left[ \frac{iv_t}{2}(\psi_R ^a\partial_x \psi_R ^a
-\psi_L ^a\partial_x \psi_L ^a) + im_t\psi_R^a \psi_L^a \right]\nonumber\\
\end{eqnarray}
where
\begin{eqnarray}
\label{masstriplet}
m_t &\approx&(J_{\bot} -2\,J_{\times})+{\rm sign}(J_{\bot} -2\,J_{\times}) \, \tilde {\gamma}  \, ,\nonumber\\
m_s &\approx&-3(J_{\bot} -2\,J_{\times})+{\rm sign}(J_{\bot} -2\,J_{\times}) \, \tilde{\gamma}
\end{eqnarray}
and
$\tilde {\gamma}\sim  \exp\left({-2\pi}/{\gamma}\right)>0$ in weak coupling. The
appearance of the $sign$ function in the expressions for the masses
of the Majorana fermions is {\it the most important
result of the mean-field treatment}.
Eq.\ (\ref{masstriplet}) shows that only the singlet among the
Majorana fermions can soften by varying interchain interactions.
Note that the triplet and singlet Majorana fermions can be
interpreted as follows in the limit of strong rung coupling:
the triplet $\psi^a$ corresponds to a
rung magnon (a triplet excitation on a rung), while the singlet
$\rho$ corresponds to the singlet bound state of two magnons.
Indeed, we will argue soon that in the
context of the effective field theory the singlet Majorana fermion softens
at Ising phase transitions. On the other hand, we will
show in section \ref{sec:BS} that the same Ising phase transitions
can be traced to the softening of a singlet bound state of two magnons in
our exact diagonalization data.

It is well-known that a
Majorana fermion describes the
long-distance properties of the two-dimensional
Ising model,\cite{Itzykson} with the mass of the fermion proportional to
the deviation from criticality $m \sim (T-T_c)/T_c$, meaning in
particular that positive mass corresponds to the disordered phase
of the Ising model.
The following local representations hold in Ising variables
for the order parameters of the columnar and staggered dimerized
phases $\epsilon_{\pm}$:\cite{Shelton,GNT}
\begin{equation}
\label{dimising}
\epsilon_+ \sim \mu_1 \mu_2 \mu_3 \mu_4 \, , \quad
\epsilon_- \sim \sigma_1\sigma_2\sigma_3\sigma_4 \, ,
\end{equation}
where $\mu_i$ and $\sigma_i$ are disorder and order variables of the $i$-{th}
replica of the two-dimensional Ising model respectively.\cite{KadanoffCeva71}

\begin{figure}[tb!]
\centering
\includegraphics[width=70mm]{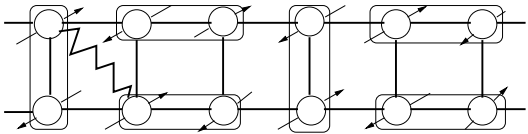}
\caption{\label{fig:rvb} Typical configuration in the RVB state of
the antiferromagnetic ladder.}
\end{figure}

{}From the above formulas we read off the following picture: if we add to
decoupled dimerized chains a relevant coupling with positive relevant exchange,
$J_{\bot} -2\, J_{\times}>0$,
then four copies of the Ising model will be in the disordered state,
meaning that the ground state will have columnar dimer order.
In the case of $J_{\bot} -2\,J_{\times}<0$, a staggered dimerized
ground state will be selected.
This analysis is supported by the following topological arguments:
the ground state of the ordinary antiferromagnetic ladder is parity symmetric,
$\left\langle \epsilon_{\pm} \right\rangle=0$. A typical configuration
in the resonating valence bond state of the antiferromagnetic ladder
is depicted in Fig.~\ref{fig:rvb}, with rectangular boxes indicating singlets
formed between the nearest-neighbor spins and the zigzag line
representing an effective $S=1$ spin formed across the ladder diagonal.

One can introduce even and odd topological (`string') order parameters,
as described in Refs.~\onlinecite{Solyom,White96}.
These string order parameters have a simple geometric
interpretation: they count the number of valence bonds crossed by
an arbitrary vertical line such that the even string order
parameter is non-zero
if the number of crossed bonds is even and the odd one is non-zero
if this number is odd.
A nice feature of the two-leg ladder systems is that
short-range valence-bond configurations have definite
parity string order parameter.\cite{Solyom}

\begin{figure}[tb!]
\centering
\includegraphics[width=60mm]{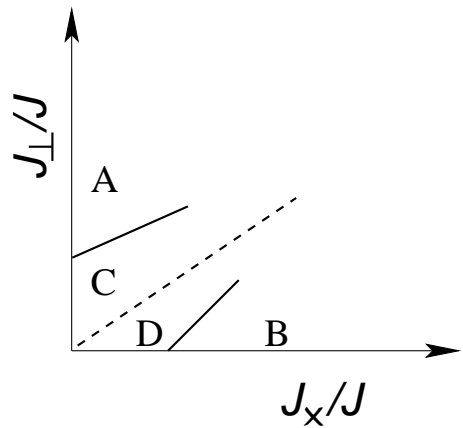}
\caption{\label{fig:weakcoupling}Topology of the phase diagram as captured
within the weak-coupling approach.
Phases are as follows: A- rung singlet, B- Haldane, C- columnar dimer and D- staggered dimer.
The dashed line represents a first-order phase transition line,
whereas continuous lines stand for Ising transition lines.
Deviations of the dashed line at weak coupling from the straight line
$J_{\bot}=2\,J_{\times}$ and enlargement of the columnar dimer phase
are due to second order-processes as found in
Ref.\ \onlinecite{Balents}.}

\end{figure}

In the rung-singlet phase the even string order parameter is non-zero.
Since $\gamma$ does not introduce coupling between the chains both patterns
of dimerization (columnar and staggered) are energetically
equally favorable at this level,
and from Fig.\ \ref{fig:rvb} it is clear that the columnar dimer
configuration will be long-range ordered after adding relevant interchain
coupling with $J_{\bot} -2 \, J_{\times}>0$.
Similar arguments apply to the case  $J_{\bot} -2 \, J_{\times}<0$.
{}From the topology of typical configurations of the RVB state one
can see in this case that singlets formed along the ladder diagonals
coexist with staggered dimer configurations.\cite{White,White96}
The string order parameter is odd and upon increasing inchain frustration
the staggered dimer phase will be selected for the ground state.
Crossing the surface which at weak couplings reads
$J_{\bot}-2\, J_{\times}\sim0$, the string order parameter jumps from even to
odd by a first-order phase transition. These findings are summarized by
the qualitative weak-coupling phase diagram Fig.~\ref{fig:weakcoupling}.

Eq.~(\ref{masstriplet}) also shows that gaps open linearly as one
deviates from the Ising transition lines either in the rung singlet
or in the Haldane phase directions.

Let us summarize the mean-field scenario presented above and
the supporting topological arguments:
when two dimerized chains are coupled in such a way that relevant
interchain coupling is present (in the model which we consider
it is always present at weak couplings\cite{Balents}) it provides a
confining potential between the gapped spinons of the individual chains
and the four-fold degeneracy of the ground state is reduced to two fold.
But what can we say when interchain coupling is itself frustrated,
i.e.\ when no relevant terms stem from interchain exchange and no
confining potential is supplied for deconfined gapped spinons
(for example, if we add additional four-spin interactions)?
We answer these issues in the next section where we derive
a new quantum phase transition when both competing interactions
(inchain as well as interchain) are marginal.

\subsection{Mapping to the Gross-Neveu model}

\label{sec:GN}

In this section we fine-tune interchain exchange to cancel relevant
contributions. As recently discussed by Starykh {\it et al.},\cite{Balents}
in the simple model (\ref{LatticeHamiltonian})
interchain exchange cannot be fine-tuned to yield only marginal
interactions unless we add four-spin counter terms.\cite{Tsvelik} In real
systems small four-spin interactions are always present e.g.\ represented
by the cyclic ring exchange. So we can assume that we start from
the line $J_{\bot}-2\,J_{\times}=0$ and relevant terms generated in
second order are compensated with small four-spin counter terms.
In that case the low-energy degrees of freedom of our fine-tuned model
will be described by the triplet and singlet massless Majorana fermions
coupled by marginal four-Fermi interactions alone:
\begin{eqnarray}
\label{Gross}
H_{\rm marg}&=&J \, a_0\int dx
\big[ (\lambda+\gamma)(\psi_R^1 \psi_L^1\psi_R^2 \psi_L^2\nonumber\\
&+&\psi_R^2 \psi_L^2\psi_R^3 \psi_L^3+\psi_R^1 \psi_L^1\psi_R^3 \psi_L^3)\\
&-&(\lambda-\gamma)(\psi_R^1 \psi_L^1+\psi_R^2 \psi_L^2+\psi_R^3 \psi_L^3)\rho_R\rho_L\big]
\, ,
\nonumber
\end{eqnarray}
where $\lambda =J_{\bot}+2\,J_{\times}=2\,J_{\bot}$.
As long as $\gamma<\lambda$, including negative $\gamma$ we can repeat
the analysis by Allen {\it et al.}\cite{Allen} where neglecting the
anisotropy a mapping to the $O(4)$ Gross-Neveu model\cite{Gross} was used.
Based on the one-loop RG equations one can check that negative $\gamma$
flows to zero at low energies and neglecting the anisotropy
will not affect the ground-state properties of the model, neither will
the excitation spectrum be modified qualitatively.
With increasing $\gamma$ to zero the mapping to the $O(4)$ Gross-Neveu
model becomes only a better approximation and exactly when $\gamma=0$
the $O(4)$ Gross-Neveu
model is the exact low-energy effective field theory describing chains coupled
by frustrated interchain interaction. Further increasing inchain frustration,
telescoping (\ref{Gross}) shows that at the point where
\begin{equation}
\label{qcp}
\gamma=\lambda=2\,J_{\bot}
\end{equation}
a new quantum phase transition takes place where the zeroth Majorana fermion
decouples from the rest of the triplet.
Low-energy excitations at this quantum critical point
are governed by a free massless Majorana fermion plus
an $O(3)$ Gross-Neveu model with dynamically generated mass for the remaining
triplet of Majorana fermions. In the extreme limit $\gamma \gg \lambda$
we can neglect the small anisotropy between triplet and
singlet Majorana fermions and recover the $O(4)$-invariant Gross-Neveu model,
with dynamically generated mass.
The sign of the mass defines the dimerization pattern of the model.
Elementary excitations are deconfined massive spinons which interpolate
between the four-fold degenerate vacua of the lattice model. This theory
describes a first-order phase transition line between the columnar dimer and
staggered dimer phases.


\section{Numerical results}

In order to test numerically the predictions for the phase diagram and the
nature of phase transitions we start
from the Majumdar-Ghosh point $J_2 = J/2$ in the decoupled
chains.\cite{Majumdar} Each of the decoupled chains then has two degenerate
dimerized ground states with a gap to the excitations. This behavior
is generic in the vicinity of the Majumdar-Ghosh point, but the
correlation length is minimal exactly at $J_2 = J/2$.\cite{WA96}
Such a short correlation length is advantageous for numerical calculations
since it minimizes finite-size effects.

For two weakly coupled dimerized chains, there will be four states
which are well separated from higher excitations. We then need
to clarify which ground state is formed depending on the coupling
between the chains. In this context, it is interesting to note that
for the particular choice $J_{\bot}=2\,J_{\times}$ of interchain
coupling at $J_2 = J/2$, {\it the two staggered dimer arrangements remain
exact eigenstates}. Indeed, it is straightforward to check that interchain
coupling terms cancel on the staggered dimer states. On the other
hand, columnar dimer states cease to be exact eigenstates of the
full ladder Hamiltonian, thus they can lower their energy by
fluctuations. Therefore, we expect that two chains with $J_2 = J/2$
weakly coupled by $J_{\bot}=2\,J_{\times}$ will have
a columnar dimer pattern in the ground state.

To determine the ground state phase diagram, we have performed exact
diagonalization of ladders with up to $N=16$ rungs (32 spins)
using a standard Lanczos method, exploiting the conservation
of total $S^z$ and spatial symmetries. We impose periodic boundary conditions
along the chains in order to have translational symmetry. Furthermore,
there is an exchange symmetry of the chains. Important information about
the ground state is provided by the associated quantum numbers,
in particular momentum $k$ for
translation along the chains, and $k_y$ for the `momentum' corresponding
to an exchange of the chains. On a finite system,
the two staggered and columnar dimer states
both combine to a ground state with $k=0$, $k_y = 0$.
However, the value of $k_y$ in the first excited state
distinguishes between the two cases: a combination of staggered dimer
states gives rise to $k=\pi$, $k_y = \pi$, while one obtains $k=\pi$,
$k_y = 0$ from the columnar dimer states.

Extrapolations to the thermodynamic limit $N \to \infty$
will be performed using the
Vanden-Broeck-Schwartz (VBS) algorithm\cite{VBS79,HeSch88}
(with the parameter $\alpha = -1$ of the algorithm).

\subsection{Low-lying levels}

\label{sec:level}

\begin{figure}[tb!]
\centering
\includegraphics[width=\columnwidth]{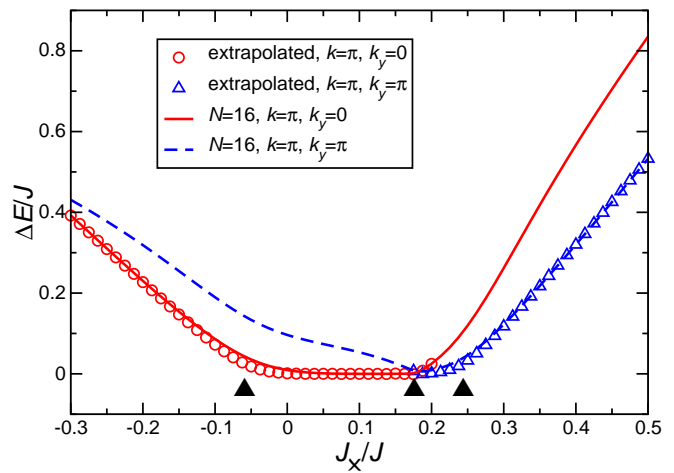}
\caption{\label{fig:Gaps} (Color online)
Numerical results for the lowest excited states with $S=0$ and $k=\pi$
at $J_2 = J/2$, $J_\bot = 0.2\,J$ as a function of $J_\times$.
Lines are for $N=16$ rungs, symbols show extrapolations to the thermodynamic
limit. The big filled triangles denote estimates for the locations
of the different phase transitions. For details compare the text.
}
\end{figure}

To illustrate the generic behavior, we will first discuss a cut
through the phase diagram at $J_\bot = 0.2\,J$. The lines in
Fig.~\ref{fig:Gaps} show the excitation energy relative to the
ground state for the lowest levels with $k=\pi$, $k_y = 0$ and
$\pi$ on a finite system
with $N=16$ rungs. On the left side of Fig.~\ref{fig:Gaps}, all
excitations are gapped, consistent with a rung-singlet phase.
For $0 \lesssim J_\times \lesssim 0.18\,J$,
the $k=\pi$, $k_y=0$ level approaches the ground state, as
is expected in a columnar dimer phase.
Note that the point $J_{\bot} = 0.2\, J$, $J_{\times} = 0.1\,J$
lies inside the columnar dimer phase, as is expected since it
belongs to the line $J_{\bot}=2\,J_{\times}$ discussed above.
On the other hand,
for $J_\times \approx 0.2\,J$, it is the level with
$k=\pi$, $k_y = \pi$ which is approaching the ground state.
This behavior is consistent with a staggered dimer phase
around $J_\times \approx 0.2\,J$. Finally, on the right
side of Fig.~\ref{fig:Gaps}, all excitations are again
gapped, consistent with a Haldane phase.

The locations of the phase transitions shown by the
three big filled triangles in Fig.~\ref{fig:Gaps} are
determined as follows. For intermediate $J_\times$,
the levels shown in Fig.~\ref{fig:Gaps} are the only
low-lying excitations, implying a direct first-order
transition from the columnar dimer to the staggered
dimer phase. We take the crossing of
these two excited levels on a ladder with $N$ rungs as an
estimate for the transition point. We then extrapolate
the crossing points with $6 \le N \le 16$ rungs using the VBS algorithm.
This extrapolation leads to
the middle triangle in Fig.~\ref{fig:Gaps}. Since the
dependence on $N$ is only weak, this extrapolated point
is very close to the finite-size estimate for $N=16.$

The transitions to the rung-singlet phase and the
Haldane phase at the left and right boundaries are
expected to be second-order phase transitions. Such
transitions can be accurately estimated using the
phenomenological renormalization group method:\cite{RW80}
The critical couplings $\{J\}$ are determined by the
condition that the levels for two different system
sizes $N_1$, $N_2$ satisfy
\begin{equation}
N_1 \, \Delta E_{k=\pi, k_y}(N_1, \{J\}) =
N_2 \, \Delta E_{k=\pi, k_y}(N_2, \{J\}) \, .
\label{eq:PRG}
\end{equation}
Here, one should choose $k_y = 0$ and $\pi$ for the
transitions into the columnar and staggered dimer phases,
respectively. Application of (\ref{eq:PRG}) to
$N_1 = 14$ and $N_2 = 16$ yields the two remaining filled
triangles in Fig.~\ref{fig:Gaps} (we neglect the
dependence on $N_1$ and $N_2$ since in this case
it is only weak).

Finally, the open circles and triangles in Fig.~\ref{fig:Gaps}
show an extrapolation to the thermodynamic
limit of the $k_y = 0$ and the $k_y = \pi$
levels, respectively. This extrapolation has
been performed with system sizes starting at
$N=4$ rungs using the VBS algorithm.
Errors of the extrapolation are difficult to estimate,
but should not exceed the size of the symbols.
In some cases, no convergence can be seen in the
finite-size data for the higher-lying level such that
no extrapolation can be performed for that level. However,
in these regions of $J_\times$,
the energy of the higher level shows a tendency to increase
with $N$ for the accessible system sizes.
Thus, the higher-lying level can safely be assumed to
stay at high energies in those regions where we cannot
extrapolate it.

The behavior of the extrapolated gaps is essentially
linear approaching the two
second-order phase transitions from the outside.
This is consistent with a critical exponent $\nu = 1$,
characteristic of the two-dimensional Ising
universality class\cite{Onsager,FB67} predicted
above.

\subsection{Phase diagram}

\begin{figure}[tb!]
\centering
\includegraphics[width=\columnwidth]{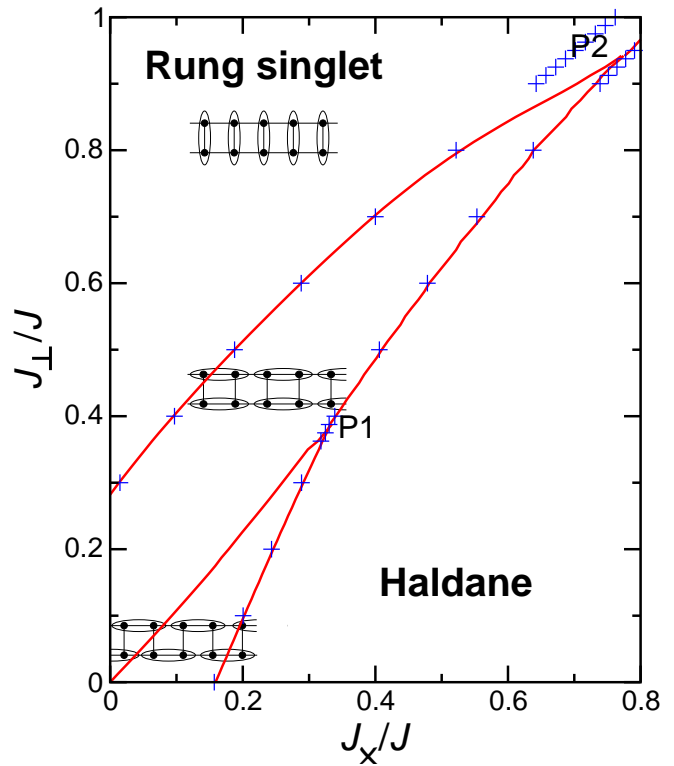}
\caption{\label{fig:Phase2D} (Color online)
Numerical results for the phase diagram in the $J_2=J/2$ plane.
Rung singlet, Haldane, columnar dimer and staggered dimer phases
can be distinguished. Crosses have been determined by
the phenomenological renormalization group condition (\ref{eq:PRG})
with $N_1=14$, $N_2=16$, the corresponding lines with
$N_1=12$, $N_2=14$. For further details and the nature of the
phase transitions compare the text.}
\end{figure}

Now we apply the results of the preceding subsection to determine
the full phase diagram for $0 \le J_\times \le J$, $0 \le J_\bot$ in the
plane $J_2 = J/2$. Numerical scans were performed for $N \le 14$ rungs
since they require diagonalizations for a large set of data points
(we have performed computations for approximately 4000 different
pairs of $J_\times$ and $J_\bot$ at $N=14$). In selected regions
we performed additional computations for $N=16$. Fig.~\ref{fig:Phase2D}
summarizes our results. One can distinguish four different gapped
phases. A rung-singlet and a Haldane phase appear at large $J_\bot$
and $J_\times$, respectively. Previous results\cite{Capriotti} for
the point $J_\times = 0$, $J_\bot = J$, $J_2 = J/2$ are consistent
with our conclusion that this lies well inside the rung-singlet phase.
The rung-singlet and the Haldane phases have a non-degenerate
ground state for a system with periodic boundary conditions.
Between these two phases there exist two dimerized phases for small
values of $J_\times$ and $J_\bot$, namely a staggered and a columnar
dimer phase. The latter phases have a two-fold degenerate ground state
with a spontaneously broken translational symmetry in the thermodynamic
limit $N \to \infty$.

The transition between the staggered and columnar dimer phase is
of first order and was determined as described in subsection \ref{sec:level}:
level crossings were first determined for fixed $6 \le N \le 14$
($16$ close to P1) and then extrapolated to $N\to \infty$
using the VBS algorithm. The error of this extrapolation is at most
on the order of the width of the line.
This transition line passes through the origin, as is expected by
consideration of two decoupled dimerized chains.

\begin{figure}[tb!]
\centering
\includegraphics[width=\columnwidth]{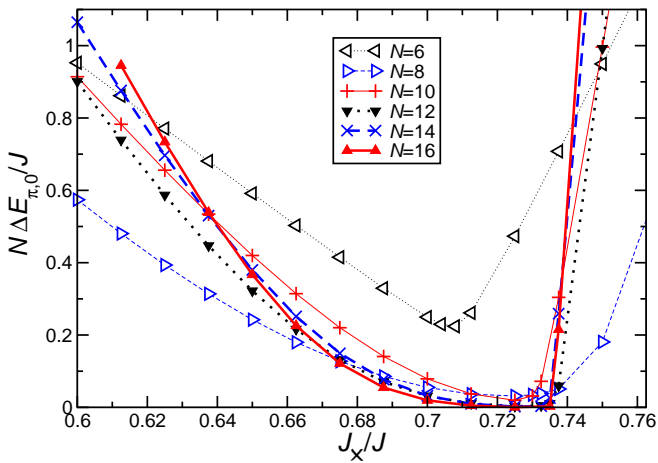}
\caption{\label{fig:ColPrg} (Color online)
Scaled energies of the $k=\pi$, $k_y = 0$ excited level
for $J_\bot = 0.9\,J$, $J_2 = J/2$ as a function of $J_\times$.
Lines connect actual numerical results which are shown by symbols.
According to the phenomenological renormalization group
condition (\ref{eq:PRG}), crossings of levels for different $N$
are estimates for the boundaries of the columnar dimer phase.
}
\end{figure}

The transitions from the dimerized phases to the rung-singlet
or Haldane phases were determined by the
phenomenological renormalization group method (\ref{eq:PRG})
with $N_1 = 12$, $N_2 = 14$ (lines) and
$N_1 = 14$, $N_2 = 16$ (crosses). This is certainly justified for
the transition between the columnar dimer and rung-singlet
phase as well as the transition between the staggered dimer
and Haldane phase, since these should be second-order transitions
belonging to the two-dimensional Ising universality class.
The transition between the columnar dimer and Haldane phase
was also estimated by
the phenomenological renormalization group method, although this
may actually be a first-order transition. Nevertheless,
finite-size effects are small (see e.g.\ right side of Fig.~\ref{fig:ColPrg}).
Therefore, our estimates for the transition
between the columnar dimer and Haldane phase are probably
quite accurate, even if we may not have used the most appropriate
method.

The intermediate dimerized phases disappear (at least according
to the $N_1=12$, $N_2=14$ estimates) for large interchain couplings
at points P1 and P2, respectively.
Beyond the point P2, one expects a direct first-order
transition from the Haldane to the rung-singlet phase, like in
the case $J_2 = 0$.\cite{Wang,Gelfand} As in Ref.~\onlinecite{Wang},
we determine the location of this first-order transition from
a cusp in the ground-state energy, using the data for $N=14$ rungs.
The transitions between the rung-singlet and Haldane phase
at $J_2 = 0$ and $J_2 = J/2$ probably belong to the same
universal surface of first-order transitions.

Finite-size effects of the transition lines are small in most
regions of  Fig.~\ref{fig:Phase2D}, as can be
seen by comparison of the results for $N_1=12$, $N_2=14$ and
$N_1=14$, $N_2=16$.
This applies in particular to the neighborhood of P1 whose location
can therefore be considered accurate.
However, close to P2, we observe finite-size
effects of the boundaries of the columnar dimer phase.
In order to discuss this in more detail, Fig.~\ref{fig:ColPrg}
shows scaled energies of the $k=\pi$, $k_y = 0$ excited level
for $J_\bot = 0.9\,J$, $J_2 = J/2$ and $6 \le N \le 16$.
The crossings between the scaled levels for different $N$
estimate the phase boundaries according to the phenomenological
renormalization group condition (\ref{eq:PRG}). Unfortunately
there is no systematic dependence of the crossings points
on $N_1$, $N_2$ such that
an extrapolation is not possible. Most crossings for the
transition to the rung singlet phase (left side of Fig.~\ref{fig:ColPrg})
lie between the $N_1=14$, $N_2=16$ crossing at $J_\times \approx 0.642\,J$
and the $N_1=12$, $N_2=14$ crossing at $J_\times \approx 0.704\,J$.
Furthermore, linear extrapolation of the levels for fixed $N$
suggests a closing of the gap and thus a phase transition around
$J_\times \approx 0.66\,J$. Overall, it seems likely that the
$N_1=12$, $N_2=14$ data underestimates the stability region of
the columnar dimer phase, while $N_1=14$, $N_2=16$ may overestimate
it. Thus, the location of the termination point P2 is probably
at larger $J_\times$, $J_\bot$ in the thermodynamic limit than
the estimate in Fig.~\ref{fig:Phase2D}. Still one can argue that the
columnar dimer phase has to terminate. Firstly, it cannot extend beyond
$J_\times = J$, since the duality transformation mentioned
at the beginning of section \ref{secWeak} maps the columnar
dimer phase which we observe for $J_\times < J$ to a different
phase with dimers on the diagonals for $J_\times' = J > J' = J_\times$,
$J_\bot' = J_\bot$. Secondly, the region where $J_\bot$ is much
bigger than all other exchange constants obviously belongs
to the rung-singlet phase.\cite{BDRS93} Therefore, the columnar dimer
phase has to terminate either at a point P2 located in the region
$J_\times < J$, or on the line $J_\times = J$. If the latter situation is
realized we cannot exclude the possibility that the rung-singlet and Haldane
phase are separated by a Gaussian quantum critical point located at $J_\times = J$.

Finally, we discuss the relation of the numerical phase diagram
to the picture obtained by the mapping to the Gross-Neveu model
presented in subsection \ref{sec:GN}.
Numerically we see that the columnar dimer phase extends to stronger
interchain couplings than the staggered dimer phase.
This gives further support to the observation that additional
interactions such as ring exchange are necessary to fine-tune
all relevant interchain couplings to zero. Indeed,
it was found that ring exchange stabilizes staggered dimer
order.\cite{Vekua,Troyer} Therefore, ring exchange will cause the
points P1 and P2 in Fig.~\ref{fig:Phase2D} to move towards each other.
P1 and P2 will merge when all relevant interchain couplings are
fine-tuned to zero. For the latter situation the effective theory
of the multi-critical point will be described by one gapless Majorana fermion
decoupled from the rest of the gapped $O(3)$ Gross-Neveu model as
presented in subsection \ref{sec:GN}.
Since the effective field theory describing long wavelength fluctuations of 
our model has a form similar to the one considered in Ref.~\onlinecite{Balents},
the same conclusion concerning the need to include additional four-spin
interactions in order to suppress all relevant interchain couplings
applies to the case $J_2=0$ considered in Ref.~\onlinecite{Balents}.

\subsection{Softening of a bound state}

\label{sec:BS}

\begin{figure}[tb!]
\centering
\includegraphics[width=\columnwidth]{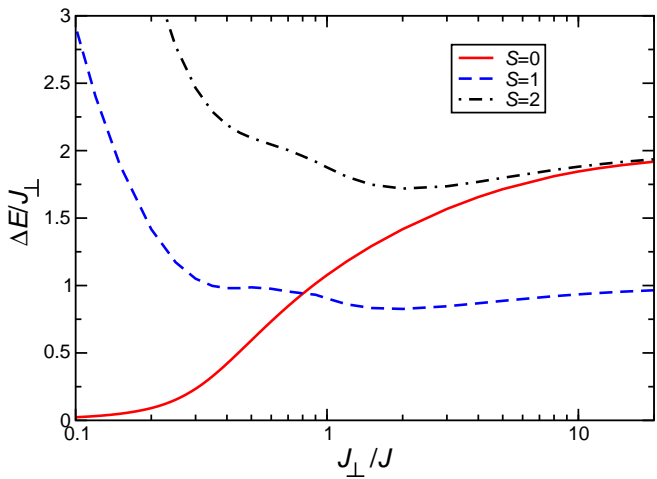}
\caption{\label{fig:Bound} (Color online)
The lowest excited levels with total spin $S=0$, $1$, and $2$ on a
ladder with $N=12$ rungs for $J_\times = 0$, $J_2 = J/2$
as a function of $J_\bot$ on a semi-logarithmic scale.
}
\end{figure}

In this last subsection we will comment on the nature of the
elementary excitations. For this purpose we start from
strong $J_\bot$ where one naturally obtains a rung-singlet
phase\cite{BDRS93} and approach the transition to the columnar dimer phase
along the line $J_\times = 0$. Since we just wish to make
a qualitative point, we restrict to $N=12$ rungs in
this section. Fig.~\ref{fig:Bound} shows
numerical results at $J_\times = 0$, $J_2 = J/2$ for the
lowest excited levels in the sectors with total spin $S=0$, $1$,
and $2$. At strong $J_\bot$ the lowest excitation is
a propagating triplet around $\Delta E = J_\bot$
above a non-degenerate ground state.
On the other hand, the singlet excitation which softens
at the transition to a columnar dimer state at $J_\bot \approx
0.29\, J$ can be traced to the lower boundary the $S=2$
continuum which starts at $\Delta E = 2\, J_\bot$ for $J_\bot \to \infty$.
This implies that the singlet which at small $J_\bot$ is a low-energy
excitation originates from a bound state of two triplets at strong
$J_\bot$. For the parameters of Fig.~\ref{fig:Bound}, this
bound state has lower energy than the fundamental triplet
for $J_\bot \lesssim 0.8\, J$ and finally softens at the
Ising transition while states with non-zero total spin
remain at a finite energy.
Similarly, the transition from the Haldane phase to the staggered
dimer phase should occur via softening of an $S=0$ two-magnon bound
state while a single magnon remains gapped at the transition.

It is straightforward to analyze the one-triplet and
the bound-state problem of two triplets by
perturbation theory at strong $J_\bot$.
In fact, formation of bound states has been observed
for ladders with $J_2 = J_\times = 0$ (see e.g.\
Ref.\ \onlinecite{SK98}). However, in this case none of the bound
states has lower energy than the lowest edge of the
two-triplet continuum. Two things happen when one turns
on $J_2$ or $J_\times$. Firstly, the minimum of the
single-triplet dispersion generally shifts to incommensurate
wave vectors. Secondly, now a bound state can appear
at strong coupling which has an energy below the lowest
edge of the two-particle continuum (see also
Ref.\ \onlinecite{KSE99} for $J_2 = 0$). However,
it is necessary to go to high orders\cite{Trebst}
to obtain quantitative agreement with numerical data.
Therefore we do not pursue this further here.

Finally, we would like to mention that
the bound-state nature of the lowest excitation leads to
unusual finite-size effects close to the phase transitions
and thus explains the difficulties with the
extrapolations encountered in the previous subsections.

\section{Conclusions}

We have studied the ground-state phase diagram and the nature of elementary
excitations of a frustrated  spin ladder. The effect of
relevant and marginal interchain exchanges has been
analyzed by an effective field theory approach.
Particular attention was paid to the interplay between two
clearly separate scales set by inchain frustration and
interchain couplings. First, a rung-singlet and a Haldane
phase are known to appear for $J_2=0$.\cite{Gelfand,Wang,Oitmaa,KSE99,Allen}
We predicted that for sufficiently large $J_2$, two dimerized phases
appear between the rung-singlet and Haldane phase. The transitions
from the dimerized phases to the rung-singlet and Haldane phases
are found to be in the two-dimensional Ising universality class,
whereas the transition between the columnar and staggered dimer
phases is of first order.

Furthermore, we have verified these predictions by exact diagonalization.
The full phase diagram has been determined in the plane $J_2 = J/2$, see
Fig.~\ref{fig:Phase2D}. The staggered dimer phase was found to terminate
first at a point P1 while the columnar dimer phase extends to stronger
interchain coupling, terminating at P2.
Beyond P2 there is a direct first-order transition from the rung-singlet
to the Haldane phase if P2 is located at $J_\times < J$,
like for $J_2 = 0$.\cite{Gelfand,Wang} Should P2 be located at
$J_\times = J$,\cite{remark} rung-singlet and Haldane 
phase might be separated by a second-order phase transition point.

The part of the phase diagram with $J_\times > J$
is obtained from the one with $J_\times < J$ by a
duality-type argument.\cite{Oitmaa}
In particular, the columnar and staggered
dimer phases map to two different dimerized
phases with dimers on the diagonals.

Finally, we have observed an
interesting phenomenon when approaching the columnar or staggered dimer
phase from the rung-singlet or Haldane phase, respectively. Namely,
a singlet bound state forms below the two-triplet scattering continuum.
This singlet bound state crosses the fundamental triplet excitation
and finally softens at the Ising phase transition.

Since all phases are gapped, they will be stable under weak
higher-dimensional coupling between the ladders. If one approaches
the $J_1-J_2-J_3$ frustrated square
lattice\cite{GSHLL,SachdevBhatt} by such inter-ladder coupling,
the rung singlet phase will be adjacent to a N\'eel phase,
and the Haldane phase will share a boundary with a phase where
spins are ordered antiferromagnetically
along one `ladder' direction and ferromagnetically
along the other direction. Our results show that in such a generalized
phase diagram there will be further dimerized phases, in particular
some with two-dimensional columnar and staggered dimer patterns.
We therefore hope that our results will also be useful to elucidate
the phase diagram of the $J_1-J_2-J_3$ frustrated spin-1/2 Heisenberg
antiferromagnet on the square lattice.\cite{GSHLL,SachdevBhatt}

\begin{acknowledgments}
This work was initiated at the Institut f\"ur The\-o\-re\-tische Physik (ITP),
University of Hannover. Both authors would like to thank the ITP for the
hospitality extended to them during this period.
We are particularly grateful to H.-J.\ Mikeska for valuable discussions.
Furthermore, A.H.\ would like to acknowledge
the Laboratoire de Physique Th\'{e}orique, ULP Strasbourg
for hospitality during the final stages of this work. Parts of the
numerical computations have been performed on COMPAQ ES45 and
SGI Altix 350 compute servers at the TU Braunschweig and
ULP Strasbourg, respectively.

\end{acknowledgments}


\begin{thebibliography}{10}
\bibitem{remark} Duality arguments show that
  P2 cannot be located in the region $J_{\times}>J$.
\end{thebibliography}
\end{document}